\documentclass{article}
\PassOptionsToPackage{numbers, sort&compress}{natbib}
\usepackage[preprint]{neurips_2024}
\usepackage[utf8]{inputenc} 
\usepackage[T1]{fontenc}    
\usepackage{hyperref}       
\usepackage{url}            
\usepackage{booktabs}       
\usepackage{amsfonts}       
\usepackage{nicefrac}       
\usepackage{microtype}      
\usepackage{xcolor}         
\usepackage{authblk}
\usepackage{pifont}
\usepackage{graphicx}
\usepackage{array}     
\usepackage{multirow}  
\usepackage{algorithm}
\usepackage{algpseudocode}
\usepackage{fontawesome}
\usepackage{tabularx}
\usepackage{booktabs}
\usepackage{enumitem}

\title{RedChronos: A Large Language Model-Based Log Analysis System for Insider Threat Detection in Enterprises}

\author[1,2,3]{Chenyu Li}
\author[3]{Zhengjia Zhu}
\author[4]{Jiyan He\textsuperscript{\faEnvelope}}
\author[1,2]{Xiu Zhang\textsuperscript{\faEnvelope}}

\affil[1]{Institute of Information Engineering, Chinese Academy of Sciences}
\affil[2]{School of Cyber Security, University of Chinese Academy of Sciences}
\affil[3]{Xiaohongshu}
\affil[4]{School of Cyber Science and Technology, University of Science and Technology of China}

\affil[ ]{\texttt{lichenyu1999@iie.ac.cn,zhuzhengjia@xiaohongshu.com\\
linmin@xiaohongshu.com,hejiyan@mail.ustc.edu.cn\textsuperscript{\faEnvelope},zhangxiu@iie.ac.cn\textsuperscript{\faEnvelope}}}

\begin{document}

\maketitle

\begin{abstract}
Internal threat detection (IDT) aims to address security threats within organizations or enterprises by identifying potential or already occurring malicious threats within vast amounts of logs.
Although organizations or enterprises have dedicated personnel responsible for reviewing these logs, it is impossible to manually examine all logs entirely.
In response to the vast number of logs, we propose a system called RedChronos, which is a Large Language Model-Based Log Analysis System. 
This system incorporates innovative improvements over previous research by employing Query-Aware Weighted Voting and a Semantic Expansion-based Genetic Algorithm with LLM-driven Mutations. 
On the public datasets CERT 4.2 and 5.2, RedChronos outperforms or matches existing approaches in terms of accuracy, precision, and detection rate. 
Moreover, RedChronos reduces the need for manual intervention in security log reviews by approximately 90\% in the Xiaohongshu Security Operation Center. 
Therefore, our RedChronos system demonstrates exceptional performance in handling IDT tasks, providing innovative solutions for these challenges. 
We believe that future research can continue to enhance the system's performance in IDT tasks while also reducing the response time to internal risk events.
\end{abstract}

\section{Introduction}
Internal threats account for nearly one-third~\cite{Homoliak2019} of all cybercrime sources in cybersecurity, and currently no company can completely resolve the issue of internal threats. 
Compared to external threats, which require exploiting software vulnerabilities and overcoming firewalls, internal threats inherently have access to certain parts of an enterprise or organization's internal systems. 
As a result, internal threats are more likely to cause significant damage.
To address internal threats, internal threat detection (IDT) is a technique that involves analyzing and uncovering potential threats from a vast amount of internal logs.

With advancements in machine learning technologies, IDT techniques are also rapidly evolving. Techniques such as deep learning~\cite{Yuan2018,Yuan2021} and unsupervised learning~\cite{Le2021} can identify more suitable features from the log data to effectively detect internal threats.
Some studies~\cite{Liu2019,LMTracker2022,Al-Mhiqani2022,Haitao2024} employ methods such as graph neural networks and transfer learning to explore suitable features in log data.
However, these methods often encounter common challenges in training models, such as the scarcity of anomalous data related to internal threats, model overfitting, limited generalization ability, and poor interpretability of the final results.
Consequently, an increasing number of researchers are beginning to explore the use of large pre-trained models to address log detection issues.
Currently, an increasing number of researchers are focusing their IDT detection studies on large language models (LLM), as the excellent generalization capabilities of these pre-trained models can efficiently accomplish log analysis tasks.
Methods based on LLM can primarily be divided into three approaches: full-sample learning~\cite{Xiao2023}, few-shot model fine-tuning~\cite{song-etal-2025-confront}, and prompt engineering~\cite{Chengyu2024}.These approaches leverage the transfer learning capabilities of LLMs to effectively handle this format of logs, opening up new avenues for processing logs in different formats.

However, LLMs may still not have fully realized their potential. For instance, strategies like Retrieval-Augmented Generation (RAG)~\cite{lewis2021retrievalaugmentedgenerationknowledgeintensivenlp} and Chain of Thought (CoT)~\cite{wei2023chainofthoughtpromptingelicitsreasoning}, which merely alter the way prompts are input, have significantly improved the performance of various solutions. 
Research~\cite{Chengyu2024,Xiao2023,Liu2024,Liu2024Interpretable} based on Large Language Models (LLMs) fully leverages the capabilities of these powerful models, employing Prompt Engineering methodologies to achieve comparatively good performance for IDT without requiring enormous model training costs.
Although these methods have addressed some internal threat issues, there likely exist more superior ways of structuring prompt engineering for LLMs to tackle more complex logs and scenarios.
Hallucinations, inability to follow instructions, heterogeneity in task proficiency across different LLMs, and the strong correlation between prompts and model performance remain significant challenges affecting the effectiveness of Large Language Models.

To address the aforementioned challenges, our research has led to the development of RedChronos (a Large Language Model-Based Log Analysis System for Insider Threat Detection in Enterprises). Our contributions are as follows:
\begin{enumerate}
    \item Our proposed RedChronos system not only demonstrates superior performance compared to existing methods on public datasets, but also helps Security Operations Center (SOC) engineers reduce over 90\% of their workload when deployed in enterprise internal networks.
    \item Our proposed RedChronos system employs Query-Aware Weighted Voting technology to overcome model hallucinations and enhance system accuracy. This technique performs weighted voting on final results based on the model's performance on IDT tasks, which significantly improves the overall system performance.
    \item Our proposed RedChronos system implements a Semantic Expansion-based Genetic Algorithm with LLM-driven Mutations technique to automatically update prompts for IDT tasks. This enables the system to continuously evolve its log processing capabilities, proving more scientific and efficient than conventional approaches relying on engineers' experience or manual prompt fine-tuning.
\end{enumerate}

The paper is organized as follows: In Section~\ref{sec:relatedWork}, we review related work; in Section~\ref{sec:systemdesign}, we present our system design; in Section~\ref{sec:expSetup}, we detail the experimental setup; in Section~\ref{sec:resAndAna}, we analyze the experimental results; and finally, in Section~\ref{sec:cfw}, we conclude the paper and discuss future research directions.

\section{Related Work}
\label{sec:relatedWork}
In this section, we will discuss related work on IDT. 
In previous research on IDT, numerous noteworthy studies have emerged, which can be broadly categorized into three approaches: detection methods based on traditional algorithms, methods based on training models, and approaches leveraging Large Language Models (LLMs).

Traditional approaches to insider threat and Advanced Persistent Threat (APT) detection have established foundational methodologies that employ a variety of techniques, including structural anomaly detection, psychological profiling, heterogeneous graph analysis, formal descriptive frameworks, and rule-based systems, each contributing distinct capabilities for identifying suspicious patterns in user behavior and network activities within organizational systems.
Brdiczka~\cite{Brdiczka2012} combines Structural Anomaly Detection and Psychological Profiling to proactively identify insider threats, validated using World of Warcraft player behavior data.
LMTracker~\cite{LMTracker2022} uses heterogeneous graphs and unsupervised learning to detect lateral movement paths in APTs, outperforming existing methods with high accuracy.
DD-GCN~\cite{Li2023Convolutional} detects insider threats by leveraging heterogeneous graphs that combine user behavior and structural relationships, achieving higher accuracy through dual-domain analysis.
Glasser~\cite{Glasser2013} describes implementing a synthetic data generation system for insider threat research, modeling organizational relationships, assets, communications, and behaviors with controlled threat scenarios.
Agrafiotis~\cite{Agrafiotis2016} proposes a framework for detecting insider threats using formal "tripwires" that identify policy violations or known attack patterns, integrated with anomaly detection systems.
Magklaras~\cite{Magklaras2012} presents ITPSL, a specialized language for systematically describing insider threats and misuse incidents, including its design, implementation, and evaluation against real-world scenarios.
Liu~\cite{Liu2019} proposes log2vec, a heterogeneous graph embedding method for insider threat detection that converts logs into graphs, generates vector representations, and clusters to identify malicious activities.
Sun~\cite{Sun2022} proposes HetGLM, a system using heterogeneous graphs and a novel GNN-based algorithm (MADR) to detect lateral movement in APT attacks without labeled data or preset thresholds.

Recent advances in insider threat detection demonstrate a shift toward sophisticated AI techniques.
Yuan~\cite{Yuan2018} proposes a novel insider threat detection approach using LSTM-CNN architecture, where LSTM captures temporal patterns in user behavior sequences and CNN classifies these patterns, achieving 0.9449 AUC without manual feature engineering.
Yuan's~\cite{Yuan2021} survey reviews deep learning applications for insider threat detection, demonstrating their superiority over traditional machine learning approaches by better handling complex, heterogeneous data despite challenges of limited labeled data and adaptive attacks.
Al-Mhiqani~\cite{Al-Mhiqani2020} provides a comprehensive survey that taxonomizes insider threat types and reviews detection approaches, analyzing real cases and machine learning techniques while identifying challenges and offering recommendations to mitigate insider threats.
Al-Mhiqani~\cite{Al-Mhiqani2022} proposes a multilayer framework that first selects optimal ML models using entropy-VIKOR methods, then implements a hybrid detection approach combining Random Forest for known threats and KNN for unknown threats, achieving 97-99\% accuracy.
Nasir~\cite{Nasir2021} proposes a deep learning approach for insider threat detection through behavioral analysis, using rich feature sets including logon events and user roles, outperforming existing techniques with 97\% precision on CERT dataset.
Le~\cite{Le2021} develops an unsupervised anomaly detection approach using four diverse methods with temporal data representations, combined through voting-based ensembles to improve performance and robustness, detecting 60\% of malicious insiders with only 0.1\% investigation budget.
Alsowail~\cite{Alsowail2021} proposes a unified multi-tiered framework integrating technical, psychological, behavioral, and cognitive factors to address insider threats throughout the complete employment lifecycle, from pre-employment screening through post-departure monitoring.
DeepLog~\cite{Du2017} applies LSTM neural networks to model system logs as natural language sequences, learning normal patterns automatically and detecting anomalies when logs deviate, with capabilities for online updating and workflow construction to aid diagnosis.
Xiao~\cite{Haitao2024} proposes CATE, a framework integrating statistical analysis via convolutional attention networks and sequential analysis via transformer encoders to detect insider threats, effectively combining both dimensions of user behavior data.
LogGPT~\cite{Xiao2023} employs GPT architecture for log anomaly detection, first training it to predict next log entries, then applying reinforcement learning fine-tuning specifically for anomaly detection, outperforming existing approaches on three datasets.
LAN~\cite{Cai2024} is a real-time activity-level insider threat detection framework that employs graph structure learning to model both temporal dependencies and cross-sequence relationships while using a hybrid prediction loss to address data imbalance challenges.
Qi~\cite{Qi2023} leverages ChatGPT's language interpretation capabilities for log-based anomaly detection through prompt engineering, transferring knowledge from large language models to analyze system logs with good performance and interpretability on benchmark datasets.

Recent advances in applying large language models to cybersecurity have produced innovative frameworks for insider threat detection and log analysis, with specialized prompting, multi-agent approaches, and privacy-preserving techniques enhancing detection capabilities.
Audit-LLM~\cite{Chengyu2024} employs a three-agent framework (Decomposer, Tool Builder, Executor) for insider threat detection, enhanced by Evidence-based Multi-agent Debate to improve accuracy and faithfulness when analyzing log files.
The method~\cite{song-etal-2025-confront} fine-tunes LLMs with a two-stage contrastive learning strategy using natural language behavior representations to precisely distinguish benign anomalies from actual insider threats.
Haywood~\cite{Haywood2025} uses LLMs to detect insider threat sentiment in job reviews, generating synthetic data to overcome ethical concerns and benchmarking LLM sentiment analysis against human expert evaluations.
Yilmaz's survey~\cite{Yilmaz_Can_2024} paper synthesizes AI techniques for insider threat detection, examining behavioral analytics, NLP, LLMs, and graph-based approaches while addressing dataset limitations and privacy challenges.
FedITD~\cite{Qiang2024} combines federated learning with parameter-efficient tuning of large language models, enabling privacy-preserving insider threat detection while customizing models for each organization without central data sharing.
Chen's review paper~\cite{CHEN2024104016} systematically examines the applications of large language models in various cyber threat detection tasks, identifying suitable use cases, optimization points, and limitations across different security contexts.
Patil's review paper~\cite{Patil2024} examines how LLMs can improve cloud security through enhanced anomaly detection, threat intelligence generation, and incident response automation, while addressing associated challenges and ethical considerations.
LogPrompt~\cite{Liu2024Interpretable} uses large language models with specialized prompting strategies for interpretable log analysis, requiring no in-domain training while outperforming traditional methods and providing human-readable explanations for detected anomalies.
Le's study~\cite{Le2023} evaluates ChatGPT's effectiveness for log parsing through different prompting methods, finding that few-shot prompting yields promising results while identifying challenges and opportunities for LLM-based log parsing.
LogPrompt~\cite{Liu2024} leverages large language models with advanced prompting techniques for zero-shot log analysis, providing interpretable results without requiring in-domain training data while maintaining performance on unseen logs.
PSE framework~\cite{yu2023leveragingpartialsymmetrymultiagent} adaptively exploits partial symmetry in multi-agent reinforcement learning, theoretically justified by bounded performance error, improving sample efficiency and performance under various symmetry-breaking conditions in both simulations and real robots.
Egil's paper~\cite{karlsen2023benchmarkinglargelanguagemodels} benchmarks and fine-tunes 60 different LLM architectures for security log analysis, creating the LLM4Sec pipeline that enables DistilRoBERTa to achieve state-of-the-art performance across diverse log datasets.

\section{System Design}
\label{sec:systemdesign}
In this section, we will present the overall system design of RedChronos.

RedChronos will accept log entries in text format, which include, but are not limited to, email logs, security software reports, browser security logs, and internal network gateway security logs. 
After processing the logs, RedChronos will ultimately update them in its own Audit System. 
If certain logs pose a significant threat to enterprise security, they will be classified and forwarded to either regular security engineers or security experts within the security operations team.

\begin{figure}[htbp]
    \centering
    \includegraphics[width=0.8\textwidth]{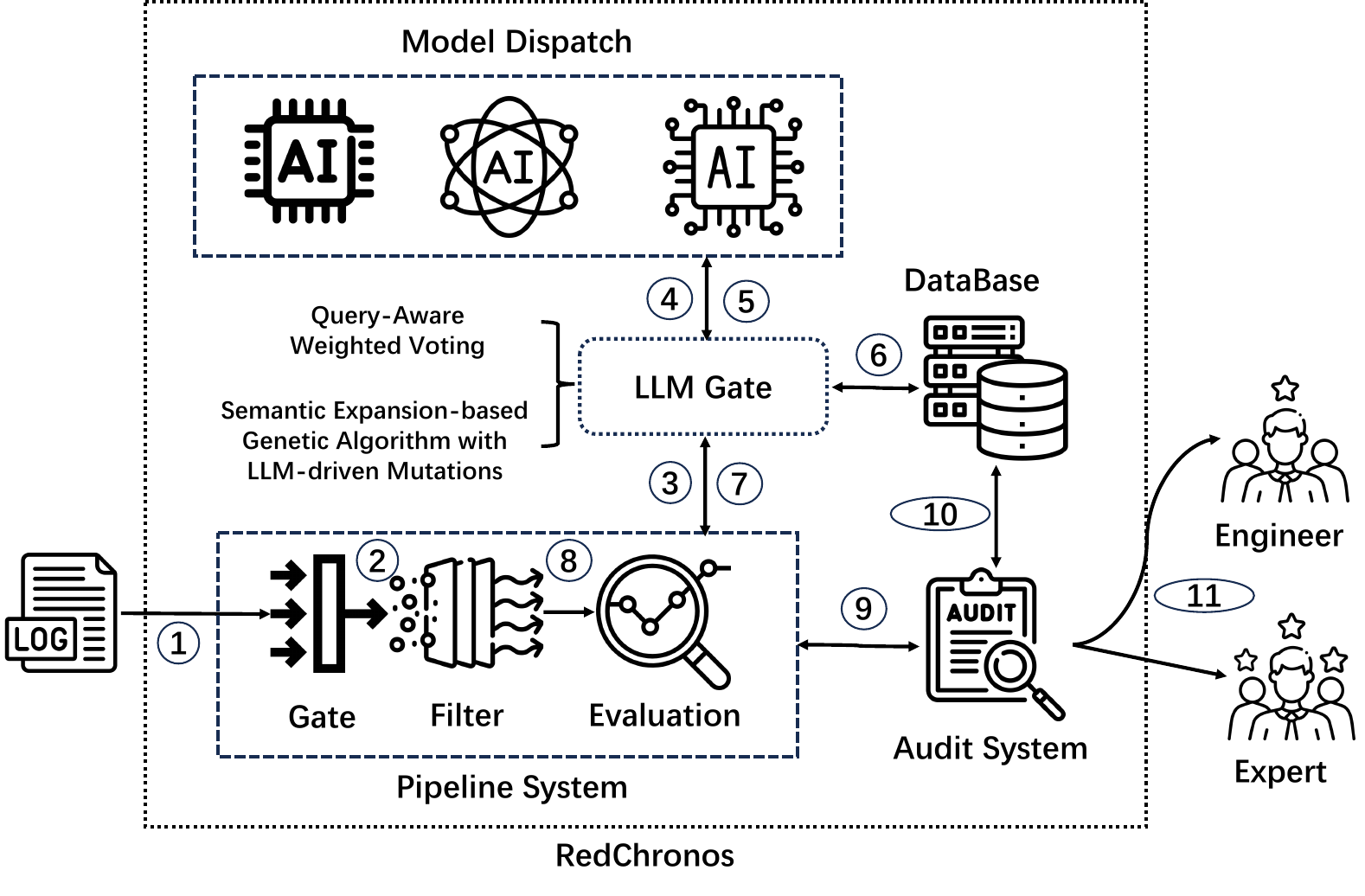}
    \caption{The overall system design of RedChronos features circles with numbers that represent the flow of information throughout the entire system when log entries are input.} 
    \label{fig:redchronosoverview} 
\end{figure}

As illustrated in Figure~\ref{fig:redchronosoverview}, the entire RedChronos consists of five subsystems: the Pipeline System, LLM Gate, Model Dispatch, Database, and Audit System.

\subsection{Pipeline System}
The Pipeline System is the first component of RedChronos that interacts with incoming logs. 
This system is primarily composed of three subsystems: Pipeline Gate, Pipeline Filter, and Pipeline Evaluation.

The Pipeline Gate serves as the sole entry point of the entire RedChronos system, responsible for handling requests sent from other systems, such as the browser security log reporting module. 
The Pipeline Gate accepts logs submitted by other systems via the HTTP protocol. 
It performs security checks, such as pattern matching and detection of potential attacks against LLM, and verifies whether the log type is supported by the RedChronos system. 
Once these checks are completed, the processed logs are forwarded to the Pipeline Filter module.

The Filter subsystem is primarily responsible for sending logs to the LLM Gate for analysis. 
It undertakes the task of secondary filtering of logs, determining based on the type of log whether it should be forwarded to the LLM Gate for further analysis. 
If the logs submitted to RedChronos originate from records processed by other internal security mechanisms of the enterprise, they are directly handled by the Evaluation subsystem. 
If there is no explicit processing status for the log, it is submitted to the LLM Gate for analysis.

The Evaluation subsystem provides a confidence score regarding RedChronos's ability to handle a given log based on predefined mathematical rules and formulas. 
This evaluation formula is designed using an engineering-based approach that considers the type of log, the performance of the LLM model processing the log, and the result of the log analysis provided by the LLM. 
The final evaluation score, along with the logs and the LLM's processing results, are then submitted to the Audit System.

\subsection{LLM Gate}
The LLM Gate is the primary analysis, processing, and scheduling module responsible for all LLM-related analyses within RedChronos, serving as a core component of the system.
LLM Gate primarily performs two tasks: the first is Query-Aware Weighted Voting, and the second is Semantic Expansion-based Genetic Algorithm with LLM-driven Mutations.

For the first task, LLM Gate comprehends the logs passed from the Pipeline System, then selects appropriate prompt templates from the Database, and sends them to the Model Dispatch module for analysis. 
Finally, LLM Gate uses the capabilities of models in the Model Dispatch to perform weighted voting on the results to derive a conclusion.

For the second task, LLM Gate evolves the prompts used for specific analytical tasks in the database. 
LLM Gate employs large language models to analyze the semantic information of existing prompts, then merges several prompts to generate a set of candidate prompts. 
For these candidate prompts, LLM Gate evaluates them using datasets. 
Through continuous repetition of this process, LLM Gate identifies the best-performing prompt, and this procedure is repeated in subsequent rounds of prompt evolution.

\begin{figure}[htbp]
    \centering 
    \includegraphics[width=0.8\textwidth]{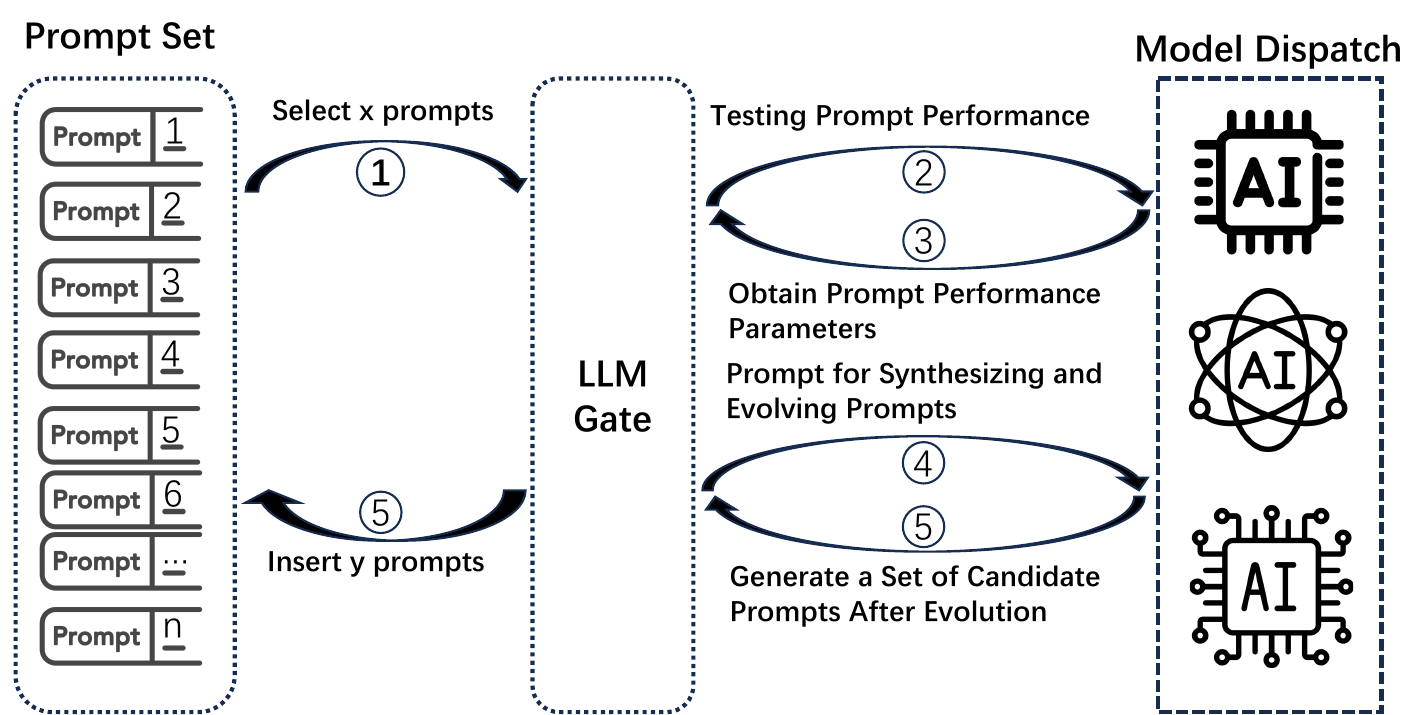}
    \caption{Flow chart of the Semantic Expansion-based Genetic Algorithm, where circles 1 to 5 represent a single execution cycle of the algorithm.} 
    \label{fig:gao} 
\end{figure}

\subsubsection{Query-Aware Weighted Voting}
Query-Aware Weighted Voting is an algorithm developed for LLM Gate that addresses performance evaluation and confidence levels of model results in log analysis tasks. 
When performing log analysis using prompts, LLM Gate's performance is primarily measured through several metrics: Accuracy (Acc), Precision (Prec), Detection Rate (DR), and False Positive Rate (FPR), all of which are used to analyze prompt effectiveness. 
For LLM Gate, these four performance parameters are predominantly influenced by two key variables: the Large Language Model itself and the content of the prompts used.

For different LLM, inherent performance variations~\cite{Al-Mhiqani2020,CHEN2024104016,Homoliak2019} exist across diverse tasks. 
Therefore, in Section~\ref{sec:resAndAna}, we conducted performance evaluations of various models using identical prompts. 
After compiling statistics on the results obtained with the same prompts, we were able to determine how different models perform on the specific task in question. 
This comparative analysis revealed the distinctive performance characteristics of each model when applied to our target application.

Different prompts significantly influence the output content of large language models. 
This variation in prompts primarily stems from two scenarios. 
First, different tasks require different categories of prompts, as illustrated in the pipeline system in Figure~\ref{fig:redchronosoverview}, where various components serve distinct purposes. 
Second, for any given task, there exists the challenge of identifying the most appropriate and effective prompt that yields optimal results. 
Regarding these two prompt-related issues, we will provide detailed explanations in Section~\ref{sec:sga}, addressing both the task-specific prompt selection and optimization methodologies for maximum effectiveness.

\begin{algorithm}
\caption{Query-Aware Weighted Voting Algorithm}
\label{alg:QAWVA}
\begin{algorithmic}[1]
\tiny
\Function{QAWVAlgorithm}{ModelWithTaskInfo, ModelCandidates, task}
    \State answerWeight = 0
    \State allWeight  = 0
    \For{each model in ModelCandidates}
        \State modelInfo = ModelWithTaskInfo[model]
        \State thisModelWeight = \Call{CalcuModelWeight}{modelInfo.Prec, modelInfo.DR, modelInfo.Acc, modelInfo.FPR}
        \If{thisModelWeight > 80}
            \State answerWeight = answerWeight + model.predict(task) $\times$ thisModelWeight
            \State allWeight = allWeight + thisModelWeight
        \EndIf
    \EndFor
    \If{answerWeight > 0.5 $\times$ allWeight}
        \Return \texttt{abnormal}
    \Else
        \Return \texttt{normal}
    \EndIf
\EndFunction
\end{algorithmic}
\end{algorithm}

\begin{algorithm}
\caption{Calculate Model Weight}
\label{alg:calcuweight}
\begin{algorithmic}[1]
\tiny
\Function{CalcuModelWeight}{Prec, DR, Acc, FPR}
    \State ModelScore $\gets (\alpha \cdot$ Prec + $\beta \cdot$ DR + $\gamma \cdot$ Acc + $\delta \cdot $ (1-FPR)) $\cdot$ 100
    \State \Return ModelWeight
\EndFunction
\end{algorithmic}
\end{algorithm}

The specific implementation of the Query-Aware Weighted Voting Algorithm is shown in Algorithm~\ref{alg:QAWVA}. 
This algorithm is named the \emph{QAWVAlgorithm} function. 
It takes \emph{ModelWithTaskInfo} as an input parameter, which contains the performance analysis of LLMs for different subtasks derived from related work. 
\emph{ModelCandidates} serves as a list of LLMs supported by Model Dispatch. 
Each eligible LLM uses a function called \emph{predict} to analyze whether the task log is malicious. 
The \emph{predict} function outputs 1 for malicious logs and 0 for normal logs.
Subsequently, \emph{QAWVAlgorithm} performs weighted calculations on all voting results, and finally determines whether the log is malicious or not.
To ensure the accuracy of the Query-Aware Weighted Voting Algorithm, we discard results from models that demonstrate poor performance on our task. 
Through our early experiments, we discovered that when a model's weight exceeds 80 points, it guarantees the correctness and rationality of the voting results.

The model's weight calculation is shown in Algorithm ~\ref{alg:calcuweight}, primarily influenced by four input parameters and four system bias parameters. 
The four input parameters are the performance metrics of the LLM on the dataset, specifically Pre , DR, Acc, and FPR, which we obtained on the test set for this task. 
The four system bias parameters are $\alpha$, $\beta$, $\gamma$, and $\delta$, which are constrained by the requirement that their sum equals 1 ($\alpha + \beta + \gamma + \delta = 1$). 
The configuration of these four parameters represents which performance aspects RedChronos emphasizes more.
The weight coefficients can be fine-tuned based on different task scenarios to optimize performance. In a balanced scenario, the weights can be set uniformly as $\alpha = \beta = \gamma = \delta = 0.25$. For tasks that demand high precision, the weights can be adjusted to $\alpha = 0.4, \beta = 0.2, \gamma = 0.2, \delta = 0.2$. Conversely, if the priority is high recall, the recommended configuration would be $\alpha = 0.2, \beta = 0.4, \gamma = 0.2, \delta = 0.2$. Finally, to achieve a low false positive rate, the coefficients should be set to $\alpha = 0.2, \beta = 0.2, \gamma = 0.2, \delta = 0.4$.

\subsubsection{Semantic Expansion-based Genetic Algorithm}
\label{sec:sga}
The Semantic Expansion-based Genetic Algorithm is a genetic algorithm we propose that analyzes the semantics of sample data and expands the original prompt based on the semantics of these samples.
Semantic Expansion-based Genetic Algorithm is demonstrated in Algorithm~\ref{alg:SEGA}.
The function \texttt{SemanticExpansionbasedGA} is designed to enhance candidate prompts using a genetic algorithm-based approach, taking \texttt{candidatePrompts} and a \texttt{model} as inputs. The algorithm systematically processes each candidate prompt by first verifying if it has not been tested (\texttt{cP.isTest == False}). For untested prompts, it evaluates them using the \texttt{model.test(cP)} method to gather necessary performance metrics. Once a prompt is tested, its weight is calculated through the \texttt{CalcuModelWeight} function, which leverages parameters such as precision (\texttt{cP.Prec}), detection rate (\texttt{cP.DR}), accuracy (\texttt{cP.Acc}), and false positive rate (\texttt{cP.FPR}). After assigning weights, the algorithm selects the prompt with the highest weight using the \texttt{selectHeightestWeightPrompt} method. The selected prompt information, denoted as \texttt{seedPromptsInfo}, undergoes semantic expansion via the \texttt{PvboSemanticExpan} function, resulting in the creation of \texttt{newSeedPrompts}. These newly expanded seed prompts constitute the output of the algorithm, poised for further application or analysis in subsequent processes.

\begin{algorithm}
\caption{Semantic Expansion-based Genetic Algorithm}
\label{alg:SEGA}
\begin{algorithmic}[1]
\Require \textit{candidatePrompts}, \textit{model}
\ForAll{cP in candidatePrompts}
    \If{cP.isTest == \textbf{false}}
        \State model.test(cP)
    \EndIf
    \State cP.Weight = \textsc{CalcuModelWeight}(cP.Prec, cP.DR, cP.Acc, cP.FPR)
\EndFor
\State seedPromptsInfo = \textsc{SelectHeightestWeightPrompt}(candidatePrompts)
\State newSeedPrompts = \textsc{PvboSemanticExpan}(seedPromptsInfo)
\State \Return newSeedPrompts
\end{algorithmic}
\end{algorithm}

In the study by Guo~\cite{guo2024connecting}, it was found that using a genetic algorithm can enable the program to efficiently find suitable prompts. 
They proposed two methods: Genetic Algorithm and Differential Evolution. 
They conducted experimental validation of their methods and concluded that employing LLM to search for prompts is a scientifically sound approach.

\begin{figure}[htbp]
    \centering 
    \includegraphics[width=0.8\textwidth]{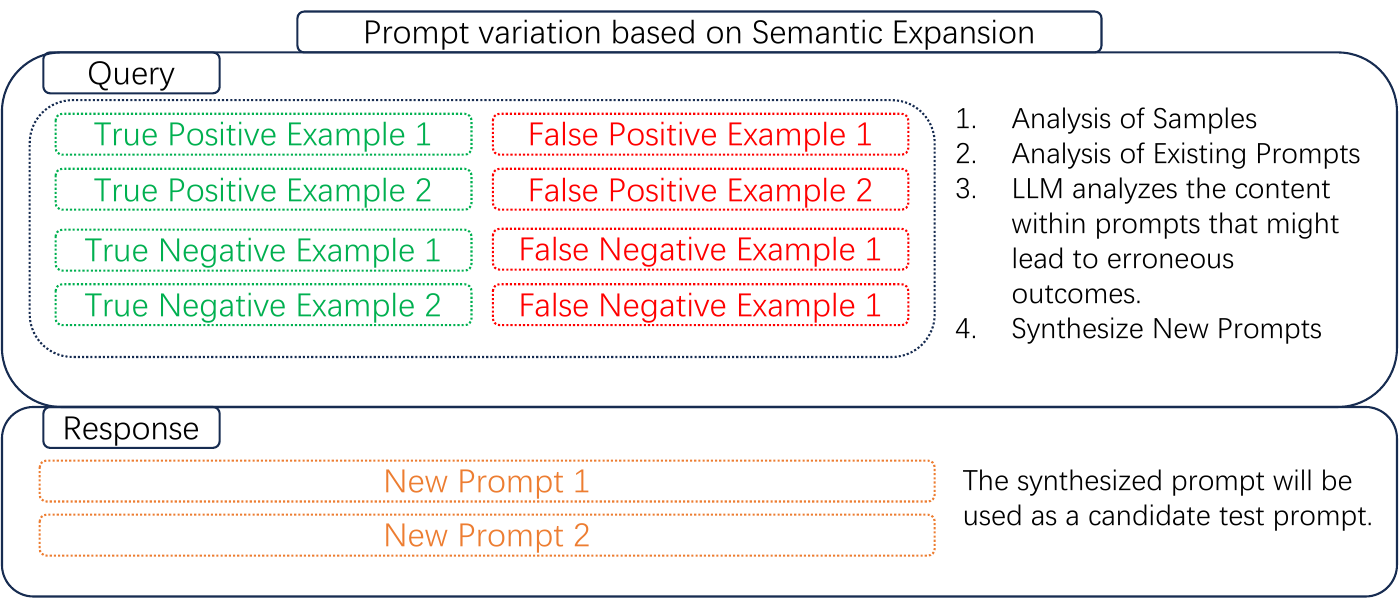}
    \caption{We use the correct and incorrect results of seed prompts on the test set as example inputs for demonstration to the LLM. An instructional prompt is then employed to guide the LLM in modifying the seed prompt based on the seed prompt itself, along with the correct and incorrect results. Finally, we utilize the new prompt as the updated seed prompt and proceed to test it. The Prompt Variation based on Semantic Expansion algorithm is denoted as PvboSemanticExpan.} 
    \label{fig:promptVariation} 
\end{figure}

Compared to existing methods, our Semantic Expansion-based Genetic Algorithm inputs the seed prompt, its performance on the test set, and specific randomly selected correct and incorrect examples as parameters for prompt variation into the LLM. 
As shown in Figure~\ref{fig:promptVariation}, the LLM generates a new prompt to serve as the seed prompt for the next round of evolution.

We employ a four-step instruction process to guide the large language model in generating new prompts according to our directives. 
In the first step, the LLM analyzes the semantics of the samples to fully understand the LLM instructions, the samples themselves, and the LLM's output. 
In the second step, the LLM examines the language of the original prompt to check for any ambiguity. 
The third step involves having the LLM focus on identifying the reasons behind incorrect examples, guiding it to modify and enhance the existing prompt rules based on this analysis. 
In the fourth step, the LLM is instructed to summarize the results from the previous three steps and synthesize a set of candidate prompts.

Since the algorithm requires us to iteratively refine the prompts and utilize a LLM to verify the effectiveness of our prompts, we also need a dataset to identify prompts with better performance. 
We selected 2000 abnormal and 2000 normal logs from the  CERT v4.2 dataset in Sec~\ref{sec:dataset}.

We begin by mutating the Seed Prompt as demonstrated in Table~\ref{tab:gademo1}. 
Subsequently, we leverage the LLM to evolve the Seed Prompt based on its performance outcomes, particularly analyzing both correct and incorrect response patterns. This iterative refinement process yields the optimized variations documented in Tables ~\ref{tab:gademo2} and ~\ref{tab:gademo3}.

The first-generation prompts in Table~\ref{tab:gademo2} demonstrate suboptimal performance metrics, which consequently leads to their exclusion from subsequent evolutionary iterations through our selection mechanism. 
In contrast, the optimized prompt variants presented in Table~\ref{tab:gademo3} exhibit superior performance characteristics. 
Through five generations of systematic evolution with elitism preservation in Table~\ref{tab:gademo4}, the final evolved prompt achieves satisfactory performance metrics.

Therefore, the Semantic Expansion-based Genetic Algorithm will automatically evolve from the seed prompt in the direction of the highest prompt score, ultimately identifying the prompt with the best performance.

\subsection{Model Dispatch}
The Model Dispatch is responsible for integrating available LLMs and providing a unified interface for them. 
These models include open-source ones such as Deepseek, Qwen, XDG, and Moonlight, as well as closed-source models like ChatGPT and Claude. 
Given the high confidentiality requirements of the internal enterprise log processing system, this paper primarily focuses on the performance when using open-source models.

\subsection{Database}
The Database is primarily used to record errors and warnings generated by RedChronos and to handle raw logs along with their processed results. 
It stores information used by various subsystems, such as Prompts and retrieval information for RAG, and provides reference information to the Audit System for SOC engineers.

\subsection{Audit System}
\label{sec:auditsystem}
The Audit System is a component developed by RedChronos to aggregate logs and process results for further analysis. 
Based on the actions suggested by the Pipeline System, the Audit System directly processes logs, such as marking certain entries in the database as handled. 
This enables the Audit System to automatically manage the majority of security incidents that pose no threat, have low threat levels, or are suitable for transactional handling. 
For security incidents that carry higher threat levels, the Audit System conducts pre-processing based on predefined protocols, including actions like revoking employee intranet access. 
These incidents are then escalated to the SOC for manual intervention. 
The Audit System categorizes the logs requiring human intervention into two levels: Security Operations Engineer and Security Expert. 
Security Operations Engineers are primarily responsible for reviewing whether RedChronos' automated log processing contains any errors, addressing events related to security updates on employee devices, and rectifying insecure configurations within internal network assets. 
Security Experts, on the other hand, handle more critical tasks such as examining potential APT (Advanced Persistent Threat) attacks, analyzing 0-Day vulnerabilities, exploiting 1-Day vulnerabilities, and conducting in-depth analysis and replication of security incidents.

\section{Experimental Setup}
\label{sec:expSetup}
In this section, we will provide a detailed description of our experimental setup, outline the experimental procedures, and conduct an analysis of the experimental results.

\subsection{Research Questions and Experimental Configuration}

\textbf{Research Questions.}
The primary research questions addressed in this experiment are listed as follows:

\begin{enumerate}
  \item \textbf{RQ1}\label{RQ1}: How does Chronos perform on public datasets compared to existing research?
  \item \textbf{RQ2}\label{RQ2}: Do different models exhibit varying performance for the same prompt?
  \item \textbf{RQ3}\label{RQ3}: Can RedChronos serve as an internal log analysis system for an enterprise or organization? How much daily workload can this system reduce?
\end{enumerate}

\textbf{Experimental Configuration.}
In this experiment, no model training process was conducted. 
All models were sourced from Xiaohongshu's internally developed ALL IN model platform, and all open-source models were deployed on Xiaohongshu's corporate intranet.

\subsection{Dataset}
\label{sec:dataset}
To validate the effectiveness of our experiments, we utilized both the publicly available Insider Threat Test Dataset CERT~\cite{Lindauer2020} dataset and the internal SOC dataset from Xiaohongshu.

The CERT dataset, released by Carnegie Mellon University, includes information such as \emph{login/logoff events}, \emph{emails}, \emph{file access}, \emph{website visits}, \emph{device usage}, and \emph{organizational structure data}. 
For our validation, we primarily used two versions of this dataset: r4.2 and r5.2.

These logs contain detailed endpoint security information, including events such as employee device logins, device usage, potential risks from outdated operating systems, the presence of suspicious software, and indications of scanning or attacks on the machines.

Due to the severe class imbalance in the CERT dataset, our experiments follow the sampling approach used in both the Unveiling Shadows~\cite{Haitao2024} and Audit-LLM~\cite{Chengyu2024} papers by restricting the number of benign class samples to below 20,000. This enables us to utilize the Audit-LLM data as our baseline for comparative analysis.

The internal SOC dataset from Xiaohongshu primarily consists of logs from various enterprise systems and security software. 
Logs from various components include email systems, gateway logs, and so on. The security software logs mainly originate from Xiaohongshu's self-developed security software, \emph{Microsoft Defender for Endpoint (MDE)}, and \emph{Eagle Cloud Technology} security logs.
We use 10,000 MDE Alert logs from September 4, 2024, to November 4, 2024, as the dataset for this paper.

\section{Results And Analysis}
\label{sec:resAndAna}

\subsection{Overall Performance}
To address \textbf{RQ}~\ref{RQ1}, we compare our approach with multiple LLM-based methods, using identical dataset parameters across all experiments. This enables a fair performance comparison between RedChronos and relevant prior work. The experimental results are presented in Table~\ref{tab:certoverall4.2} and Table~\ref{tab:certoverall5.2}.

\begin{table}[h]
    \centering
    \caption{Comparative evaluation of RedChronos against seven alternative methods on CERT r4.2 dataset. First-place results are marked in bold, and second-place results are identified with underlining. An upward arrow ($\uparrow$) represents metrics where higher values indicate better performance, while a downward arrow ($\downarrow$) represents metrics where lower values indicate better performance.}
    
    \begin{tabular}{lcccc}
        \hline
        \multicolumn{1}{c}{\textbf{Model}} & \multicolumn{4}{c}{\textbf{CERT r4.2}} \\ 
        \cmidrule(r){2-5}
        \multicolumn{1}{c}{} & \textbf{Prec $\uparrow$} & \textbf{DR $\uparrow$} & \textbf{FPR $\downarrow$} & \textbf{Acc $\uparrow$} \\ 
        \hline
        DeepLog & 0.684 & 0.715 & 0.335 & 0.743 \\
        LMTracker & 0.782 & 0.829 & 0.217 & 0.856 \\
        CATE & 0.885 & 0.892 & 0.289 & 0.928 \\
        LAN & 0.876 & 0.886 & 0.142 & 0.934 \\
        LogPrompt & 0.861 & 0.875 & 0.324 & 0.841 \\
        LogGPT & 0.911 & 0.914 & 0.184 & 0.926 \\ 
        Audit-LLM & \textbf{0.943} & \underline{0.958} & \underline{0.037} & \underline{0.961} \\ 
        \hline 
        \textbf{RedChronos} & \underline{0.933} & \textbf{0.987} & \textbf{0.022} & \textbf{0.979} \\ 
        \hline
    \end{tabular}
    \label{tab:certoverall4.2}
\end{table}

\begin{table}[h]
    \centering
    \caption{Comparative evaluation of RedChronos against seven alternative methods on CERT r5.2 dataset.}
    \begin{tabular}{lcccc}
        \hline
        \multicolumn{1}{c}{\textbf{Model}} & \multicolumn{4}{c}{\textbf{CERT r5.2}} \\ 
        \cmidrule(r){2-5}
        \multicolumn{1}{c}{} & \textbf{Prec $\uparrow$} & \textbf{DR $\uparrow$} & \textbf{FPR $\downarrow$} & \textbf{Acc $\uparrow$} \\ 
        \hline
        DeepLog & 0.728 & 0.776 & 0.264 & 0.801 \\
        LMTracker & 0.765                    & 0.794                  & 0.216                     & 0.821 \\
        CATE & 0.893                    & 0.906                  & 0.324                     & 0.932 \\
        LAN & 0.883                    & 0.891                  & 0.099               & 0.902 \\
        LogPrompt & 0.852                    & 0.862                  & 0.328                     & 0.873 \\
        LogGPT & 0.905              & 0.907          & 0.116                     & 0.918 \\ 
        Audit-LLM & \underline{0.941}           & \underline{0.956}         & \underline{0.039}            & \underline{0.959} \\ 
        \hline 
        \textbf{RedChronos} & \textbf{0.975} & \textbf{0.993} & \textbf{0.019} &  \textbf{0.988} \\ 
        \hline
    \end{tabular}
    \label{tab:certoverall5.2}
\end{table}

As shown in Table~\ref{tab:certoverall4.2} and Table ~\ref{tab:certoverall5.2}, RedChronos outperforms the state-of-the-art approach AuditLLM on public datasets in terms of Prec, DR, FPR, and Acc. 
On the public datasets CERT 4.2 and 5.2, RedChronos outperforms or matches existing approaches in terms of accuracy, precision, and detection rate.

\textbf{Superior DR and Acc.} 
RedChronos achieves the highest DR (0.987 on r4.2, 0.993 on r5.2) and Acc (0.979 on r4.2, 0.988 on r5.2) across both datasets.
This demonstrates its exceptional ability to identify true threats while maintaining high overall correctness.

\textbf{Balanced Prec and FPR.}
While Audit-LLM marginally surpasses RedChronos in Precision on r4.2 (0.943 vs. 0.933), RedChronos achieves significantly lower FPR (0.022 vs. 0.037 on r4.2; 0.019 vs. 0.039 on r5.2). 
This highlights its robustness in minimizing false alarms—a critical factor for enterprise SOC efficiency.

\textbf{Practical Relevance for SOC Workflows}
The extremely low FPR (0.022) and high Acc (0.979) across both datasets indicate RedChronos' potential to drastically reduce manual review workloads (as claimed in Section 5.3) while ensuring reliable threat prioritization.

\subsection{Ablation Study of Different Models Using the Same Prompt}
To address \textbf{RQ}~\ref{RQ2}, we conducted an ablation study using different models with the same prompt. This experiment aims to verify whether there are performance differences among various models when applied to the same task with the identical prompt.

\begin{table}[h]
    \centering
    \caption{On the CERT r4.2 dataset, the performance of various LLM models was evaluated using the same prompt.}
    \begin{tabular}{lcccc}
        \hline
        \multicolumn{1}{c}{\textbf{Model}} & \multicolumn{4}{c}{\textbf{CERT r4.2}} \\ 
        \cmidrule(r){2-5}
        \multicolumn{1}{c}{} & \textbf{Prec $\uparrow$} & \textbf{DR $\uparrow$} & \textbf{FPR $\downarrow$} & \textbf{Acc $\uparrow$} \\ 
        \hline
        DeepSeek-R1~\cite{DeepSeek-R1} & 0.667 & 0.989 & 0.160 & 0.879 \\
        DeepSeek-R1-Distill-Qwen-1.5B & 0.464 & 0.729 & 0.270 & 0.729 \\ 
        DeepSeek-R1-Distill-Qwen-7B & 0.414 & 0.937 & 0.426 & 0.662 \\
        DeepSeek-R1-Distill-Qwen-32B & 0.662 & 0.991 & 0.163 & 0.876 \\
        Llama-3.3-70B-Instruct~\cite{Llama3} & 0.332 & 0.991 & 0.644 & 0.512 \\
        DeepSeek-V3~\cite{DeepSeek-V3} & 0.933 & 0.987 & 0.022 & 0.979 \\
        \hline 
    \end{tabular}
    \label{tab:samepromotwithmodels}
\end{table}

\begin{table}[h]
    \centering
    \caption{On the CERT r5.2 dataset, the performance of various LLM models was evaluated using the same prompt.}
    \begin{tabular}{lcccc}
        \hline
        \multicolumn{1}{c}{\textbf{Model}} & \multicolumn{4}{c}{\textbf{CERT r5.2}} \\ 
        \cmidrule(r){2-5}
        \multicolumn{1}{c}{} & \textbf{Prec $\uparrow$} & \textbf{DR $\uparrow$} & \textbf{FPR $\downarrow$} & \textbf{Acc $\uparrow$} \\ 
        \hline
        DeepSeek-R1 & 0.847 & 0.804 & 0.112 & 0.850 \\
        DeepSeek-R1-Distill-Qwen-1.5B & 0.852 & 0.673 & 0.091 & 0.805 \\ 
        DeepSeek-R1-Distill-Qwen-7B & 0.815 & 0.997 & 0.175 & 0.900 \\
        DeepSeek-R1-Distill-Qwen-32B & 0.975 & 0.993 & 0.019 & 0.988 \\
        Llama-3.3-70B & 0.552 & 0.999 & 0.631 & 0.645 \\
        DeepSeek-V3 & 0.949 & 0.985 & 0.041 & 0.971 \\
        \hline 
    \end{tabular}
    \label{tab:samepromotwithmodels5.2}
\end{table}

As shown in Table~\ref{tab:samepromotwithmodels} and ~\ref{tab:samepromotwithmodels5.2}, the performance of different LLM models is significantly impacted by using the same prompt for the final results. 
Tables~\ref{tab:samepromotwithmodels} and ~\ref{tab:samepromotwithmodels5.2} present a comparative evaluation of RedChronos against seven state-of-the-art methods on the CERT r4.2 and r5.2 datasets, respectively.

\textbf{Model-Specific Performance Variability.}
Tables~\ref{tab:samepromotwithmodels} and ~\ref{tab:samepromotwithmodels5.2} reveal significant performance disparities across LLMs when using identical prompts. 
For instance, on CERT r4.2 (Table ~\ref{tab:samepromotwithmodels}), DeepSeek-V3 achieves near-perfect detection rate (DR: 0.987) and accuracy (Acc: 0.979) with minimal false positives (FPR: 0.022), while Llama-3.3-70B-Instruct exhibits catastrophic FPR (0.644) despite high DR (0.991). 
This highlights the critical impact of model architecture and training strategies on task-specific performance. Smaller distilled models (e.g., Qwen-1.5B/7B) underperform in precision (0.464) and FPR (0.270), suggesting distillation sacrifices discriminative capability for efficiency.

\textbf{Dataset-Specific Adaptation.}
Performance trends differ across CERT r4.2 and r5.2. For example, DeepSeek-R1-Distill-Qwen-32B shows improved precision on r5.2 (0.975 vs. 0.662 on r4.2), indicating better adaptation to r5.2’s threat patterns. 
Conversely, DeepSeek-R1 suffers a precision drop (0.667 to 0.847) but achieves higher DR (0.989 to 0.804) when transitioning from r4.2 to r5.2, implying a trade-off between specificity and sensitivity across datasets.

\textbf{Hallucination vs. Robustness Trade-offs}
Models like Llama-3.3-70B achieve near-perfect DR (0.999 on r5.2) but at the cost of extreme FPR (0.631), reflecting hallucination tendencies. 
In contrast, DeepSeek-V3 maintains balanced performance (DR: 0.987–0.993, FPR: 0.019–0.022), demonstrating robustness against false alarms. 
This underscores the necessity of Query-Aware Weighted Voting to mitigate unreliable model outputs.

\textbf{Scaling Benefits with Caveats.}
While larger models (e.g., Qwen-32B vs. Qwen-7B) generally improve precision and FPR, scaling alone is insufficient. 
For example, Qwen-32B outperforms its 7B counterpart in FPR (0.163 vs. 0.426 on r4.2), yet still lags behind DeepSeek-V3, emphasizing the need for specialized optimization (e.g., semantic-aware genetic algorithms) beyond mere parameter scaling.

Tables~\ref{tab:samepromotwithmodels} and ~\ref{tab:samepromotwithmodels5.2} underscore that LLM performance in IDT tasks is highly model- and dataset-dependent. RedChronos’s innovations in prompt optimization and ensemble voting address these variabilities, enabling reliable, low-intervention threat detection. Future work should explore dynamic model weighting based on real-time task feedback to further enhance adaptability.

\subsection{Research on the Application of RedChronos in SOC}
To address RQ~\ref{RQ3}, we have deployed RedChronos in Xiaohongshu’s SOC to assist security engineers and experts in log analysis. 
Based on historical log data in the SOC, RedChronos can reduce the number of log sample analyses performed by security engineers and experts by approximately 90\%.

\section{Conclusion And Future Work}
\label{sec:cfw}
Internal threat detection continues to be a challenging task in large enterprises or organizations today. 
Identifying internal threats within the vast array of logs is a critical objective. 
In this study, we propose a system called RedChronos, which employs a Query-Aware Weighted Voting mechanism and a Semantic Expansion-based Genetic Algorithm with LLM-driven Mutations. This ensures that RedChronos not only performs exceptionally well on public datasets but also addresses the instability of LLM model outputs and their performance dependency on prompts. 
Our research offers a new approach for automatically and efficiently identifying prompts in future internal threat detection tasks. 
Additionally, it provides Query-Aware Weighted Voting to support SOC personnel in log analysis within enterprises. With the aid of LLMs, engineers will be able to focus on investigating log samples that potentially pose higher threats in the future.

\bibliographystyle{plain}
\bibliography{myrefs}

\newpage
\section{Appendix}
\subsection{About This Paper}
We currently have several ongoing experiments, and we will continue to update the data and figures in the paper. The data presented in this paper represent a portion of the results from our research efforts.

\subsection{Experiments}
\begin{table}[h]
    \centering
    \caption{Instance of a Semantic Expansion-based Genetic Algorithm (Seed Prompt).}
    \label{tab:gademo1}
    \begin{tabularx}{\linewidth}{Xcccc}
        \toprule
        \textbf{Prompt} & \multicolumn{4}{c}{\textbf{CERT r4.2}} \\ 
        \cmidrule(r){2-5}
        & \textbf{Prec $\uparrow$} & \textbf{DR $\uparrow$} & \textbf{FPR $\downarrow$} & \textbf{Acc $\uparrow$} \\ 
        \midrule
        
                \textbf{Seed Prompt:} For a corporate security system, a large number of security logs are submitted daily, but most may not require manual intervention. Each log is classified into two processing statuses: malicious (ab) and benign (no). We adhere to the following workflow principles (perform step-by-step reasoning to ensure logical consistency; malicious classification takes precedence): 
        
        \textbf{Malicious Rules (ab):} 
        \begin{itemize}[leftmargin=*,nosep]
            \item ab1. If the webpage content is harmful to the company, classify as malicious
        \end{itemize}
        
        \textbf{Benign Rules (no):} 
        \begin{itemize}[leftmargin=*,nosep]
            \item no1. If input is not a security log, mark as benign
            \item no2. If no conditions met, mark as benign
        \end{itemize}
        
        \textbf{Key principles:} 
        \begin{itemize}[leftmargin=*,nosep]
            \item Any log satisfying one "ab" rule is malicious
            \item Malicious classification has higher priority
        \end{itemize} 
        & 0.9671 & 0.8809 & 0.0096 & 0.9637 \\
        \bottomrule
    \end{tabularx}
\end{table}

\begin{table}[h]
    \centering
    \caption{Instance of a Semantic Expansion-based Genetic Algorithm (Gen1 Prompt N1).}
    \label{tab:gademo2}
    \begin{tabularx}{\linewidth}{Xcccc}
        \toprule
        \textbf{Prompt} & \multicolumn{4}{c}{\textbf{CERT r4.2}} \\ 
        \cmidrule(r){2-5}
        & \textbf{Prec $\uparrow$} & \textbf{DR $\uparrow$} & \textbf{FPR $\downarrow$} & \textbf{Acc $\uparrow$} \\ 
        \midrule
        
                \textbf{Gen1 Prompt N1:} For a corporate security system, numerous security logs are submitted daily, though most may not require manual intervention. Each log is classified into two processing statuses: malicious (ab) and benign (no). We adhere to the following workflow principles (perform step-by-step reasoning to ensure logical consistency; malicious classification takes precedence):

\textbf{Malicious Rules (ab):}
\begin{itemize}[leftmargin=*,nosep]
    \item ab1. If webpage content is harmful to the company, classify as malicious
    \item ab2. If log contains sensitive keywords (e.g., "top-secret", "confidential", "covert"), classify as malicious
    \item ab3. If log involves external recruitment websites (e.g., "aol.com", "simplyhired.com", "monster.com"), classify as malicious
    \item ab4. If log contains external email addresses (e.g., "raytheon.com", "dtaa.com"), classify as malicious
\end{itemize}

\textbf{Benign Rules (no):}
\begin{itemize}[leftmargin=*,nosep]
    \item no1. If input is not a security log, mark as benign
    \item no2. If no conditions are met, mark as benign
\end{itemize}

\textbf{Key principles:}
\begin{itemize}[leftmargin=*,nosep]
    \item Immediate malicious classification upon satisfying any "ab" rule
    \item Malicious classification has priority over benign
\end{itemize}
        & 0.4216 & 0.9585 & 0.4227 & 0.6700 \\
        \bottomrule
    \end{tabularx}
\end{table}

\begin{table}[h]
    \centering
    \caption{Instance of a Semantic Expansion-based Genetic Algorithm (Gen1 Prompt N2).}
    \label{tab:gademo3}
    \begin{tabularx}{\linewidth}{Xcccc}
        \toprule
        \textbf{Prompt} & \multicolumn{4}{c}{\textbf{CERT r4.2}} \\ 
        \cmidrule(r){2-5}
        & \textbf{Prec $\uparrow$} & \textbf{DR $\uparrow$} & \textbf{FPR $\downarrow$} & \textbf{Acc $\uparrow$} \\ 
        \midrule
        
        \textbf{Gen1 Prompt N2:} For enterprise security systems, numerous security logs are submitted daily, though most may not require manual intervention. Each log is classified into two processing statuses: malicious (ab) and benign (no). The following workflow principles must be strictly followed (perform step-by-step reasoning to ensure logical consistency; malicious classification has priority):

\textbf{Malicious Rules (ab):}
\begin{itemize}[leftmargin=*,nosep]
    \item ab1. If webpage content is harmful to the company, classify as malicious
    \item ab2. If log contains sensitive keywords (e.g., "top-secret", "confidential", "covert"), classify as malicious
    \item ab3. If log references external recruitment websites (e.g., "aol.com", "simplyhired.com", "monster.com", "indeed.com"), classify as malicious
    \item ab4. If log contains external email addresses (e.g., "raytheon.com", "dtaa.com"), classify as malicious
\end{itemize}

\textbf{Benign Rules (no):}
\begin{itemize}[leftmargin=*,nosep]
    \item no1. If the input is not a security log, mark as benign
    \item no2. If no conditions are satisfied, mark as benign
\end{itemize}

\textbf{Core Principles:}
\begin{itemize}[leftmargin=*,nosep]
    \item Immediate malicious classification upon triggering any ab rule
    \item Malicious classification takes precedence over benign
    \item Logical reasoning must be explicitly demonstrated
\end{itemize}
        & 0.9744 & 0.8509 & 0.0071 & 0.9583 \\ 
        \bottomrule
    \end{tabularx}
\end{table}

\begin{table}[h]
    \centering
    \caption{Instance of a Semantic Expansion-based Genetic Algorithm (Gen5 Prompt N1).}
    \label{tab:gademo4}
    \begin{tabularx}{\linewidth}{Xcccc}
        \toprule
        \textbf{Prompt} & \multicolumn{4}{c}{\textbf{CERT r4.2}} \\ 
        \cmidrule(r){2-5}
        & \textbf{Prec $\uparrow$} & \textbf{DR $\uparrow$} & \textbf{FPR $\downarrow$} & \textbf{Acc $\uparrow$} \\ 
        \midrule
        
        \textbf{Gen5 Prompt N1:} For enterprise security systems, numerous security logs are submitted daily, though most may not require manual intervention. Each log is classified into two processing statuses: malicious (ab) and benign (no). The following workflow principles must be strictly followed (perform step-by-step reasoning to ensure logical consistency; malicious classification has priority):
        \textbf{Malicious Rules (ab):}
\begin{itemize}[leftmargin=*,nosep]
\item ab1. If webpage content is harmful to the company, classify as malicious
\item ab2. If log contains sensitive keywords (e.g., "top-secret", "confidential", "covert"), classify as malicious
\item ab3. If log references external recruitment websites (e.g., "aol.com", "simplyhired.com", "monster.com", "indeed.com"), classify as malicious
\item ab4. If log contains job-related keywords (e.g., "job", "resume", "salary", "hiring") that explicitly indicate recruitment or job-seeking intent in the context, classify as malicious
\item ab5. If log contains technical terms unrelated to company operations (e.g., "engineer", "recruiter", "relocation") that explicitly indicate irrelevance to company business in the context, classify as malicious
\item ab6. If log contains multiple career-development keywords (e.g., "growth", "skills", "experience", "management") that explicitly indicate professional advancement intent in the context, classify as malicious
\item ab7. If log contains multiple business-irrelevant keywords (e.g., "sales", "customer", "industry", "platform") that explicitly indicate irrelevance to company operations in the context, classify as malicious
\end{itemize}

\textbf{Benign Rules (no):}
\begin{itemize}[leftmargin=*,nosep]
\item no1. If the input is not a security log, mark as benign
\item no2. If no conditions are satisfied, mark as benign
\end{itemize}

\textbf{Core Principles:}
\begin{itemize}[leftmargin=*,nosep]
\item Immediate malicious classification upon triggering any ab rule
\item Malicious classification takes precedence over benign
\item Logical reasoning must be explicitly demonstrated
\end{itemize} 
        & 0.9370 & 0.9831 & 0.0212 & 0.9798 \\ 
        \bottomrule
    \end{tabularx}
\end{table}

\end{document}